\documentclass[aps,pre,twocolumn,showpacs]{revtex4}

\usepackage{epsfig}
\usepackage{graphicx}
\usepackage{amsmath}

\begin{document}

\title{Solutions of Fisher and Burgers' equations with finite
transport memory}

\author{Sandip Kar} 

\author{Suman~Kumar~Banik}
 
\author{Deb~Shankar~Ray}

\email{pcdsr@mahendra.iacs.res.in}

\affiliation{Indian Association for the Cultivation of Science, Jadavpur,
Calcutta 700 032, India}

\date{\today}

\begin{abstract}
The Fisher and Burgers' equations with finite memory transport, describing 
reaction-diffusion and convection-diffusion processes, respectively have 
recently attracted a lot of attention in the context of chemical kinetics, 
mathematical biology and turbulence. We show here that they admit exact 
solutions. While the speed of the traveling wavefront is dependent on the 
relaxation time in Fisher equation, memory effects significantly smoothen 
out the shock wave nature of Burgers' solution, without making any influence 
on the corresponding wave speed. We numerically analyze the ansatz for the 
exact solution and show that for the reaction-diffusion system the strength
of the reaction term 
must be moderate enough not to exceed a critical limit to allow
travelling wave solution to exist for appreciable finite memory effect.
\end{abstract}

\pacs{87.10.+e, 87.15.Vv, 87.23.Cc, 05.45.-a}

\maketitle

%\section{introduction}

\noindent
{\it Introduction}:
A number of nonlinear phenomena in physical \cite{ld}, chemical \cite{ire}
and biological processes \cite{jdm,nfb}
are described by the interplay of reaction and diffusion or by the interaction
between convection and diffusion. The well-known partial differential 
equations which govern a wide variety of them are Fisher \cite{raf}
and Burgers' \cite{jmb}
equations, respectively. While the Fisher equation describes the dynamics of 
a field variable subject to spatial diffusion and logistic growth, Burgers'
equation provides the simplest nonlinear model for turbulence. Since spatial
diffusion is common to all these processes, Fick's law forms the key element
in the description of transport. This description however, gets significantly
modified when the memory effects are taken into account, i.e., when the 
dispersal of the particles are not mutually independent. This implies that the
correlation between the successive movement of the diffusing particles may be
understood as a delay in the flux for a given concentration gradient. Over the
last several years the analysis of memory effects in diffusive processes have
attracted a lot of attention 
\cite{fort,gr1,gr2,eeh,kph1,kph2,th,vm1,wh1,wh2,mhk,ga,sf,jms,vm4,jf,kbr}
in chemical kinetics, mathematical biology and 
allied areas. The focal theme lies in the interesting traveling wave front
solutions which have been studied extensively by several authors under 
various approximations. The object of the present paper is to show that the
Fisher equation and the Burgers' equation with finite memory 
transport admit exact solutions. We numerically clarify the nature of the 
ansatz wherever necessary and analyze the physical implications of 
the solutions modified by relaxation effects and the related issues.

%\section{Fisher and Burgers' equations with finite memory transport}

\noindent
{\it Fisher and Burgers' equations with finite memory transport}:
The starting point of our analysis is the Cattaneo's modification \cite{cc}
of Fick's law in the form: 

\begin{equation}
\label{eq1}
J(x,t+ \tau)= -D \frac{\partial u(x,t)}{\partial x}
\end{equation}

\noindent
which takes care of adjustment of a concentration gradient at time $t$ with 
a flux $J(x,t+ \tau)$ at a later time $(t+ \tau)$, $\tau$ being the delay time
of the particles in adopting one definite direction of propagation. Here 
$u(x,t)$ denotes the field variable and $D$ is the diffusion 
coefficient of the particles.

The population balance equation for the particles on the other hand takes 
into
account of the conservation equation supplemented by a source function 
$k f(u)$ for the particles in the form 

\begin{equation}
\label{eq2}
\frac{\partial u(x,t)}{\partial t}= -\frac{\partial J}{\partial x} + k f(u)
\end{equation}

\noindent
The Fisher source function $f(u) = u (1 - u/K)$ has been the subject wide 
interest in various context. Here the first term in $f(u)$ signifies the
linear growth followed by a nonlinear decay due to the second one; $k$ and
$K$ being the growth rate constant of the population and carrying capacity
of the environment, respectively. In what follows we shall be considered with
two specific cases of the flux-gradient relation(1) for
Fisher and Burgers' problem.

%\subsection{The Fisher equation with nonlinear damping and finite
%transport memory}

\noindent
{\it A.The Fisher equation with nonlinear damping and finite
transport memory}:
We start with an expansion of $J$ in Eq.(1) \cite{mkot} upto first order in 
$\tau$ to obtain 

\begin{equation}
\label{eq3}
\tau \frac{\partial J(x,t)}{\partial t} + J(x,t) = 
- D \frac{\partial u}{\partial x}
\end{equation}

\noindent
Here $u(x,t)$ represents the density function. Differentiating (3) with 
respect to $x$ and differentiating (2) with respect to $t$ and eliminating 
$J$ from the resulting equations one has

\begin{equation}
\label{eq4}
\frac{{\partial}^2 u}{\partial t^2} + [ \beta - k f'(u)] 
\frac{\partial u}{\partial t} = \beta k f(u) 
 + w^2 \frac{{\partial}^2 u}{\partial x^2}
\end{equation}

\noindent
where we have used the following abbreviations

\begin{equation}
\label{eq5}
 \beta = 1/\tau  \:\:\:\: {\rm and} \:\:\:\:
 w^2 = \beta D
\end{equation}

\noindent
Eq.(4), a hyperbolic reaction - diffusion equation is a generalization of
Fisher equation for finite memory transport and nonlinear damping. 
It reduces to standard Fisher equation for
$\tau$=0. Over the
years the equation has drawn wide interest in the context of 
traveling wave solutions in various problems
\cite{fort,gr1,gr2,eeh,kph1,kph2,th,vm1,wh1,wh2,mhk,ga,sf,jms,vm4,jf,kbr}.
For example, Gallay and Raugel \cite{gr1,gr2}
have studied the propagation of front solution without the nonlinear
term $k f'(u)$. Horsthemke has discussed some
related issues in the problem of transport-driven instabilities \cite{wh2}.

We now look for the traveling wave solutions of Eq.(4) of the form
$u(x,t)= K U(x-ct) \equiv K U(z)$ with $z= x-ct$, where $c>0$ is the speed of 
the nonlinear wave (which, in general, is different from the linear wave $w$
dictated by the medium subject to the boundary condition:

\begin{equation}
\label{eq6}
U(-\infty)=1  \:\:\:\:\: {\rm and} \:\:\:\:\: U(+\infty)=0 \; \; .
\end{equation}

\noindent
Eq.(4) therefore after some algebra assumes the following form

\begin{eqnarray}
& & \frac{{\partial}^2 U}{\partial z^2} + [c(n - A)]
\frac{\partial U}{\partial z} - 2 A c U \frac{\partial U}{\partial z}
\nonumber \\
& & + n A m U(1 - U) = 0 = L(U)   \:\:\:\: (\rm say)
\label{eq7}
\end{eqnarray}

\noindent
where

\begin{equation}
\label{eq8}
m= w^2 - c^2 \:\:\:\: {\rm and} \:\:\:\: n= \beta / m \; \; .
{\rm and} \:\:\:\: A= k/m 
\end{equation}

\noindent
Following Murray \cite{jdm} we now introduce the ansatz,
\begin{equation}
\label{eq9}
U(z)= \frac{1}{\left [1 + a \exp(b z) \right]^s }
\end{equation}

\noindent
as a solution to Eq.(7), where $a$, $b$ and $s$ are positive constants to be
determined. Using (9) in (7) we obtain after some algebra 

\begin{eqnarray}
& &[s(s + 1) a^2 b^2 + n A m a^2 - s[c(n - A)]a^2 b 
\nonumber \\
& & - s a^2 b^2 ] e^{2 b z} + \left [ 2 a A m n - s a b^2 
- s[c(n - A)]a b \right ] e^{b z} 
\nonumber \\
& &+ n A m - 2 A c s a b e^{b z} (1 + a e^{b z})^{-s+1} 
\nonumber \\
& & -n A m (1 + a e^{b z})^{-s+2} = 0 = L(U) \; \ .
\label{eq10}
\end{eqnarray}

\noindent
Now for $L(U)=0$ for all $z$, the coefficients of $e^0$, $e^{b z}$, $e^{2 b z}
$ and $e^{3 b z}$ within the curly brackets must vanish identically. This
implies that $s$=0, 1 or 2. $s$=0 is not a possible solution since $s$ is a
positive constant by our starting assumption. For $s$=1 the coefficients of 
$e^{b z}$ and $e^{2 b z}$ of Eq.(10) yield the following relations,

\begin{equation}
\label{eq11}
s(s + 1)b^2 + n A m - s[c (n - A)] b - s b^2 = 0
\end{equation}

\begin{equation}
\label{eq12}
n A m - s b^2 - s[c (n - A)]b - 2 A c s b = 0
\end{equation}

\noindent
which can be solved to give $b$=0 and $b=-2 A c s/(s + 1)$.

Again since by initial assumption $b$ is a positive constant, both the values
of $b$ are unacceptable and $s$=1 is not a correct choice.

For $s$=2 Eq.(10) reduces to a form in which the coefficient of $e^{b z}$, $e^
{2 b z}$ and $e^{3 b z}$ must satisfy the following relations

\begin{equation}
\label{eq13}
s(s + 1) b^2 + 3 n A m -2 s[c(n - A)] b - 2 s b^2 = 0 \; ,
\end{equation}

\begin{equation}
\label{eq14}
s(s + 1) b^2 + n A m - s[c(n - A)] b - s b^2 = 0 \; {\rm and}
\end{equation}

\begin{equation}
\label{eq15}
2 n A m - s b^2 - s[c(n - A)] b - 2 A c s b = 0 \; .
\end{equation}

\noindent
From Eq.(13)-(15) we obtain

\begin{equation}
\label{eq16}
%b^2 = \frac{n A m}{ s(s + 1)}
b^2 = n A m/[s(s + 1)]
\end{equation}

\noindent
and putting $n= \beta/m$, $A= k/m$ from (8) and $s$=2 in (16) we have

\begin{equation}
\label{eq17}
%b^2 = \frac{\beta k}{6 m} \; \; .
b^2 = \beta k/(6 m) \; \; .
\end{equation}

\noindent
Making use of (17) in (14) we obtain $b$ in terms of $c$ as follows.

\begin{equation}
\label{eq18}
%c = \frac{5 k \beta }{6 b (\beta - k)} 
c = 5 k \beta / [6 b (\beta - k)] 
\end{equation}

\begin{equation}
\label{eq19}
%b = \frac{5}{6 c (\frac{1}{k} - \frac{1}{\beta})}
b = 5/\left [6 c (\frac{1}{k} - \frac{1}{\beta}) \right ]
\end{equation}

\noindent
The exact speed $c$ of the traveling wave can be calculated from (17) and 
(19) using $m= w^2 - c^2$ as

\begin{equation}
\label{eq20}
c = \frac{\sqrt{\beta D}}{\left [ 1 + \frac{6}{25} (y-1/y)^2 \right ]^{1/2} }
\end{equation}

\noindent
with $y = \sqrt{\frac{\beta}{k}}$.
It may be noted that the exact value of $c$ thus derived is always greater
than $c_{min}$ where

\begin{equation}
\label{eq21}
c_{min} = \frac{w}{\left [ 1 + \frac{1}{4} (y-1/y)^2 \right ]^{1/2}} \; \; .
\end{equation}

\noindent
Again in the diffusive limit, i.e., $1/\beta \rightarrow 0$ or
$1/y \rightarrow 0$ the expression (20) results in exact Fisher value of 
$c$ as $c = 5 \sqrt{kD}/\sqrt{6}$. It is necessary to stress that this is 
not the speed selected by the front ($c = 2 \sqrt{kD}$), but it yields
$2.04 \sqrt{kD}$ which is very close to the selected value \cite{jdm}.

\begin{figure}[h]
\vspace*{-2.5cm}
\includegraphics[width=0.7\linewidth]{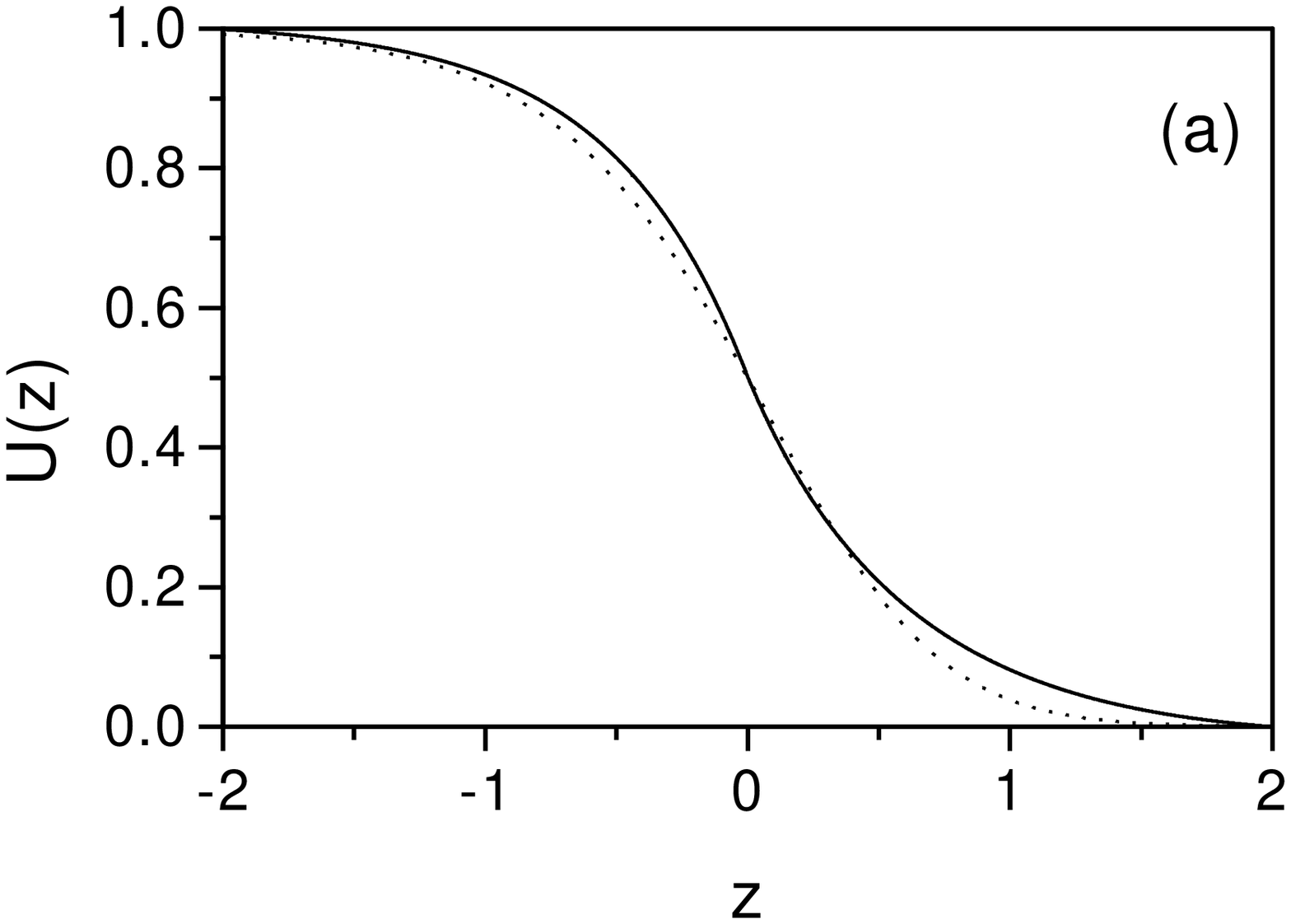}
\vspace{-2.5cm}
\vspace*{-2.5cm}
\includegraphics[width=0.7\linewidth]{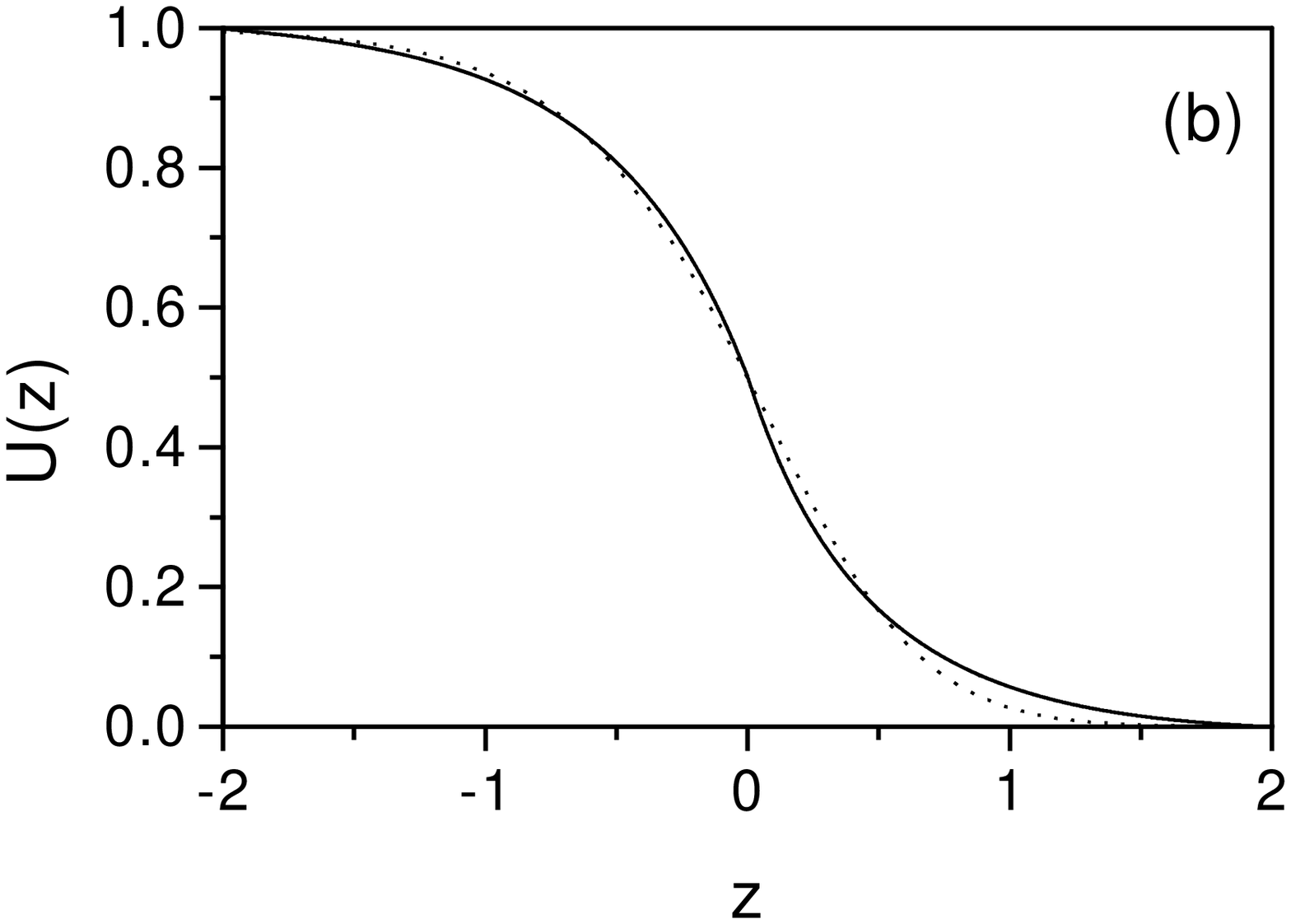}
\vspace{-2.5cm}
\vspace*{-2.5cm}
\includegraphics[width=0.7\linewidth]{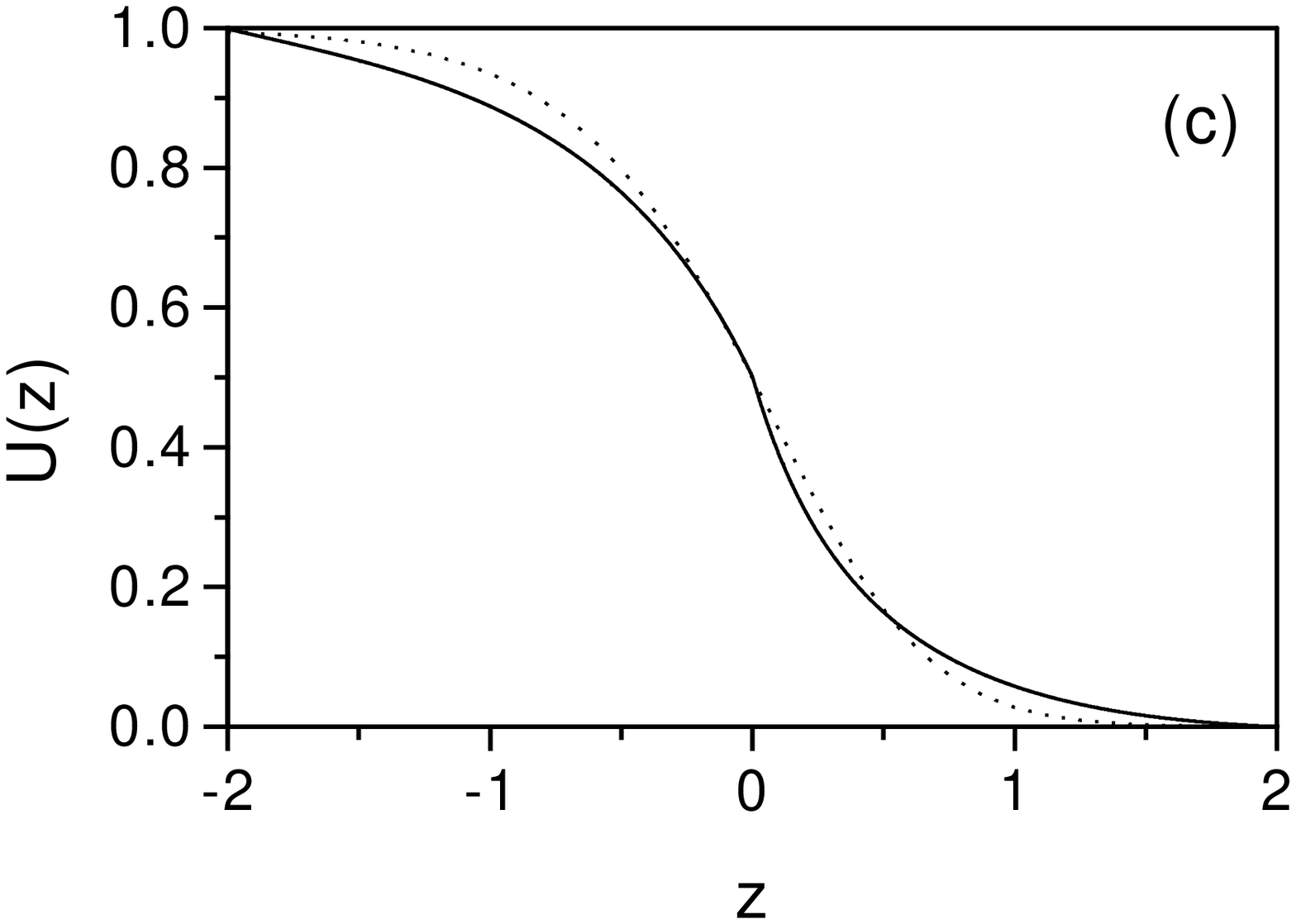}
\vspace{-2.5cm}
\vspace*{-2.5cm}
\includegraphics[width=0.7\linewidth]{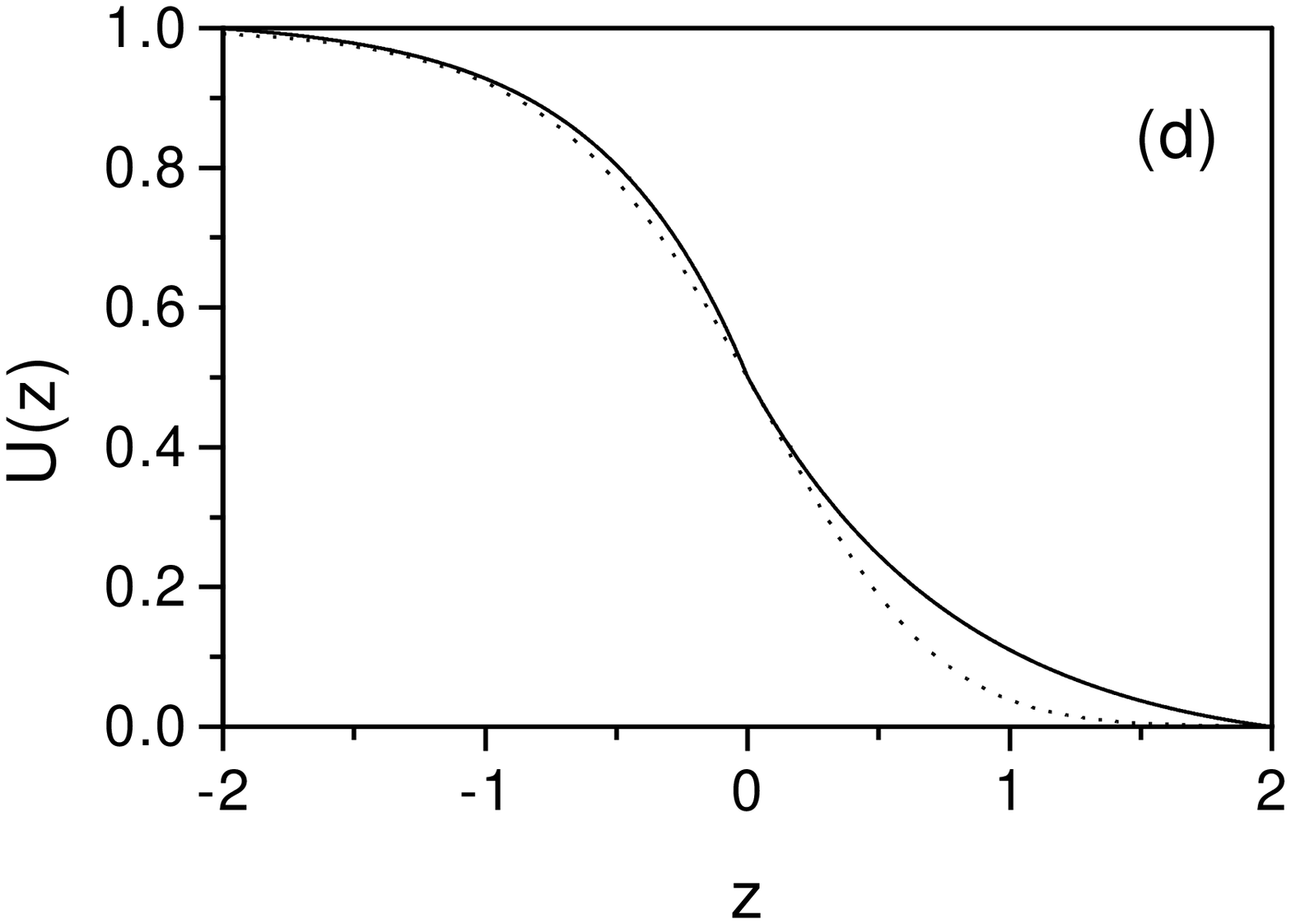}
\vspace{-2.3cm}
\caption{A plot of trvelling wave solutions for different values of  
relaxation time $\tau (= \frac{1}{\beta})$ for
$k = 0.6$ and $D = 1.0$. The solid lines are due to numerical
simulation of Eq.(4) and the dotted lines are the analytic results
(22). (a) $\tau = 0.2$, (b) $\tau = 0.4$ (c) $\tau = 0.6$ and (d) $\tau =0.0$ 
(Units are arbitrary). }
\label{fig1}
\end{figure}

Having determined $b$ and $s$ one can write down the exact form of the
traveling wave solution (9) for the problem

\begin{equation}
\label{eq22}
U(z) = 1/
\left \{ 1 + a \exp \left [ \left (
\frac{5}{c \sqrt{6} (\frac{1}{k} - \frac{1}{\beta})} \right )
\frac{z}{\sqrt{6}} \right ] \right \}^2 \; .
\end{equation}

\noindent
Furthermore $a$ can be determined from the usual condition 
$U(z) = 1/2$ for $z=0$. This results in $a= (\sqrt{2} - 1)$.
The exact solution of the Fisher equation can be recovered from (22) in the
limit $1/\beta \rightarrow 0$ (i.e., $1/y \rightarrow 0$
) using Fisher value of $c = 5 \sqrt{k D}/ \sqrt{6}$. This is given by 

\begin{equation}
\label{eq23}
U(z) = \frac{1}{ \left [ 1 + (\sqrt{2} - 1) \exp \left ( 
\sqrt{\frac{k}{D}} \frac{z}{ \sqrt{6}}\right ) \right ]^2} \; \; .
\end{equation}

We thus observe that the effect of memory or finite relaxation time enters 
into the dynamics of the reaction-diffusion system through its influence on 
the speed of the traveling wave front $c$. We emphasize here that for
$\frac{1}{\beta} = 0$ Eq.(22) does not give the solution selected by the
front but is much steeper although the speed is very close to the selected      
one. 

It is pertinent to point out that although exact the travelling wave solution
(22) does not exhaust the possibility of other solutions. This was noted
earlier by Murray \cite{jdm} in the context of fisher equation without 
memory effect which is a parabolic differential equation. For an understanding
of the nature of the travelling wave solution where $\beta = (1/ \tau)$ is a
new element of the present theory, we carry out a numerical investigation
of Eq.(4) using finite difference method to solve the boundary value problem.
The initial condition to integrate numerically is that the front is at rest
at $t = 0$. We fix the value of diffusion coefficient $D = 1.0$ for the 
entire treatment. In order to allow the variation of $\tau$ for a fixed 
value of $k$, we have kept $k$ at 0.6. For a higher value of $k$, i.e, 
where the reaction term dominates $\tau$ must be chosen appropriately over
a range to generate numerically stable travelling wave front solution. The
interplay of $\beta$ and $k$ will be considered in more detail in the later
part of this section.

In Fig.1 we compare the analytical (dotted) and the numerical (solid)
solutions corresponding to (22) and (4), respectively for different values
of $\tau$. From our analysis it is apparent that they agree fairly well for
$\tau$ roughly in the range between 0.1 and 0.5. In Fig.1(d) we present the
result for $\tau=0$, which corresponds to the typical Fisher case. The 
analytical curve is marginally steeper than numerical one. In Fig.2 we 
compare the speed of the travelling wave front computed numerically from (4)
with that obtained analytically following (20) for several values of $\tau$.
It follows that they agree reasonably well when $\tau \leq k$, i.e, in the 
range 0.1-0.5. As $\tau$ approaches zero the analytical value of $c$ becomes
lower than the numerical one. This implies that the analytical wave front
although moves slower is steeper than the numercal one since steepness goes
as $ \sim \frac{k}{c}$ as noted by Murray in his earlier analysis. For 
higher values of $\tau$ the disagreement between analytical and numerical 
values of $c$ grows rapidly.

\begin{figure}
\vspace*{-2.5cm}
\includegraphics[width=0.7\linewidth]{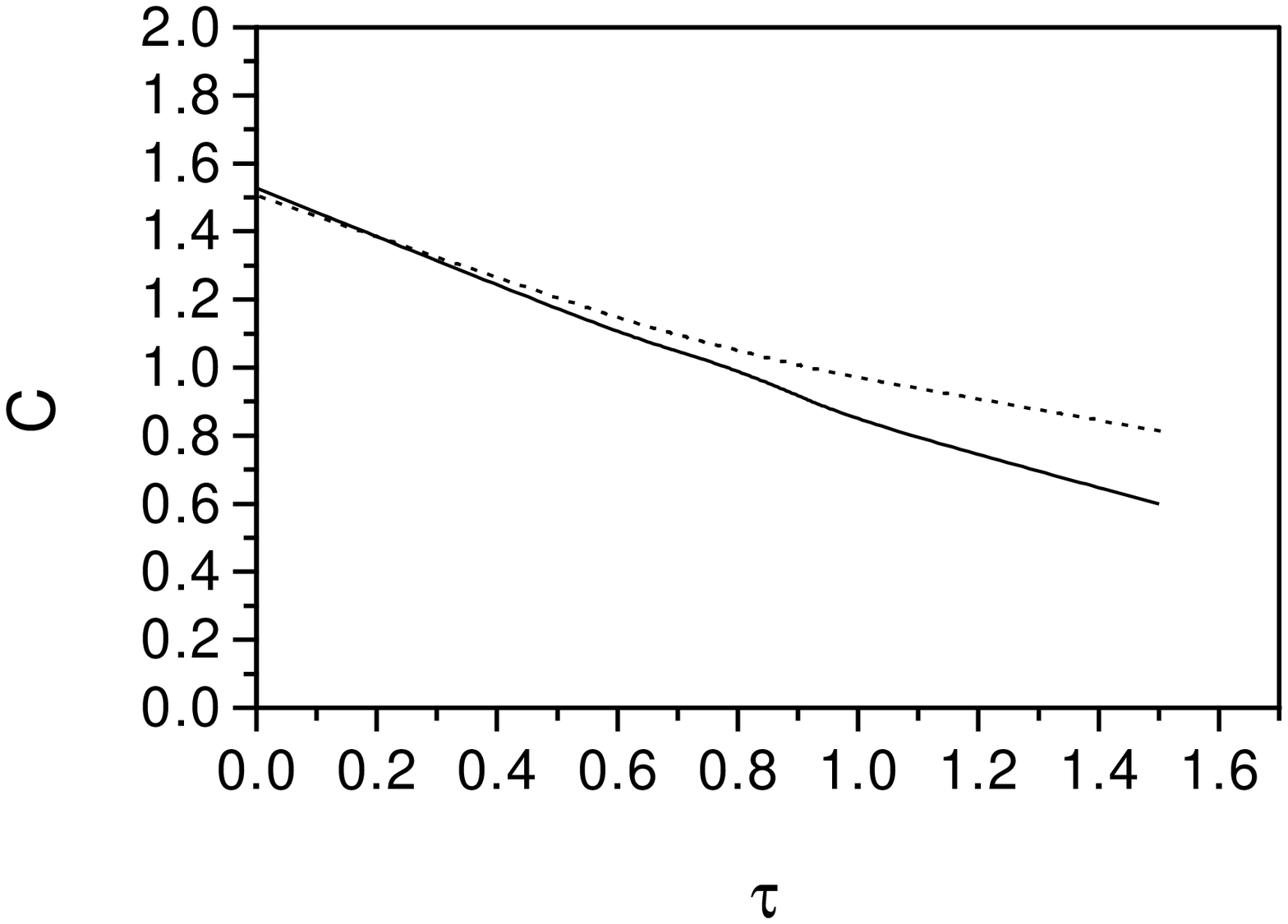}
\vspace{-2.3cm}
\caption{A plot of the speed $c$ of the travelling wave front solution vs
relaxation time $\tau$ (analytical, dotted line; numerical, solid line)
for $D = 1.0$, $k = 0.6$ (Units are arbitrary). }
\label{fig2}
\end{figure}

The above analysis suggests that there is a strong interplay of
$k$ and $\tau$ (or $\beta$) in the dynamics so far as the form and stability
of the travelling wave front solution is concerned. To explore this aspect
more clearly we now carry out an asymptotic analysis of the problem. To this
end we return to Eq.(7) subject to boundary condition (6). Following Murray
we choose the perturbation parameter $\epsilon = 1/c^2$ and look for the 
asymptotic solution for $0 < \epsilon << 1$ by introducing a change of 
variable $\xi = \frac{z}{c} = {\epsilon}^{1/2} z$ and $U(z) = g(\xi)$. With
these transformations Eq.(7) and (6) therefore reduces to

\begin{equation}
\label{eq24}
\epsilon \frac{d^2 g}{d {\xi}^2} + (n - A + 2 A g) \frac{dg}{d \xi}
+ m n A g (1 - g) = 0
\end{equation}

\noindent
and

\begin{equation}
\label{eq25}
g(- \infty) = 1 \:\:\:\:\:\: {\rm ;}   \:\:\:\:\:\:\:  g(+ \infty) = 0 \:\: .
\end{equation}

\noindent
respectively. $\epsilon$ in the highest derivative in Eq.(24) identifies it as
a singular perturbation problem.

Making use of a regular perturbation series in $\epsilon$

\begin{equation}
\label{eq26}
g(\xi ; \epsilon) = g_0(\xi) + \epsilon g_1(\xi) + ....
\end{equation}

\noindent
in (24) we obtain after equating the appropriate powers of $\epsilon$

\begin{equation}
\label{eq27}
(n - A + 2 A g_0) \frac{dg_0}{d \xi} = 
- m n A g_0 (1 - g_0) \:\:\:\:\:\:\:\:\:\:\:\:\:\:\:\: {\rm ;O(1)} \:\: .
\end{equation}

\noindent
and

\begin{equation}
\label{eq28}
(n - A + 2 A g_0) \frac{dg_1}{d \xi} + \frac{d^2 g_0}{d {\xi}^2} +
2 A g_1 \frac{dg_0}{d \xi} 
+ m n A g_1 (1 - 2 g_0) = 0 \:\:\:\:\:\:\:\:\:\:\:\:\: {\rm ;O(\epsilon)}\:\:
\end{equation}

\noindent
The lowest order equation (27) when integrated yields 

\begin{equation}
\label{eq29}
\ln \left \{\frac{(g_0)^{\beta-k}}{(1 - g_0)^{\beta+k}} \right \} =
- \beta k \xi + \beta k l
\end{equation}

\noindent
where $l$ is a constant of integration. Since we are interested in the 
solution in the vicinity of $z = 0$, i.e, $\xi = 0$ for which we put 
$g_0(\xi) = 1/2$, we obtain

\begin{equation}
\label{eq30}
l = \frac{1}{\beta k} \ln \left \{\frac{
(\frac{1}{2})^{\beta-k}}{(\frac{1}{2})^{\beta+k}} \right \}
\end{equation}

\noindent
Eq.(29) precludes the possibility of an explicit solution for $g_0(\xi)$.
Depending on $\beta$ and $k$ we therefore consider three different cases;

(i) $\beta>>k$ (or $\tau<<k$):
We have from (30) $l = 0$ and (29) reduces to 

\begin{eqnarray}
& & g_0(\xi) = (1 + \exp(k \xi))^{-1}
\nonumber \\ 
or, \:\:\:\:\:\:  & & U(z) = (1 + \exp(k z/c))^{-1} + O(\epsilon)
\label{eq31}
\end{eqnarray}

\noindent
This is the standard assymptotic solution for $U(z)$ for which the effect
of memory is negligible.

(ii) $\beta \approx k$ (i,e $\tau \approx k$):
We obtain similarly  from (29) and (30)

\begin{eqnarray}
& & g_0(\xi) = \left (1 - \frac{\exp(k \xi/2)}{2}
\right )
\nonumber \\ 
or, \:\:\:\:\:\:  & & U(z) = \left (1 - \frac{\exp((k z)/(2 c))}{2}
 \right ) + O(\epsilon)
\label{eq32}
\end{eqnarray}

\noindent
When both $\beta$ and $k$ are small compared to 1 and the exponential term
in (32) is small it is easy to put approximately the $O(1)$ term in the
form of (31) as 

\begin{equation}
\label{eq33}
U(z) \approx \left (1 + \frac{\exp((k z)/(2 c))}{2}
\right )^{-1}         
\end{equation}

(iii) $\beta<<k$ (i,e $\tau>>k$):
We obtain 

\begin{equation}
\label{eq34}
g_0(\xi) = \frac{1 \pm \sqrt{1 - 
\exp(\beta \xi)}}{2} + O(\epsilon)
\end{equation}

\noindent
The form of this solution is generically different from those of (32) and
(31) since it is independent of $k$.

The three cases discussed above clearly shows that monotonic solutions 
satisfying $U(- \infty) = 1$ and $U(\infty) = 0$ for finite wave speed 
$(c \ge c_{min})$ exist for the cases (i) and (ii), i,e , when $\tau$ is
short but finite; $\tau \le k$. The assertion of this asymptotic analysis is 
in clear agreement with our numerical simulation and our choice of a smaller
value of $k$ as discussed earlier.

The aforesaid analysis clearly demonstrates that
although the nature of the partial differential equation changes from 
parabolic to hyperbolic type due to the inclusion of relaxation time, the 
Fisher equation can be solved by Murray's ansatz \cite{jdm} to derive the 
exact wave speed and the traveling wave front solution for a suitable range 
of relaxation time $\tau$ allowed by the strength of the reaction term. 
A compromise between the exact and the numerical solution can be obtained
for relatively small reaction terms. The method can be extended further to
study other density dependent diffusive processes.

%\subsection{Burgers' equation with finite memory transport}

\noindent
{\it Burgers' equation with finite memory transport}:
The Burgers' equation \cite{jmb}
is a simple model of turbulence which illustrates an
interaction between convection and diffusion. The convection incorporates
nonlinearity in the dynamics. To include finite memory effect we proceed as 
follows:

We start with the following functional relation between flux $J(x,t+ \tau)$
at a time $t + \tau$ and the field variable $u(x,t)$ and its gradient term 
at an earlier time $t$;

\begin{equation}
\label{eq35}
J(x,t+ \tau) = \frac{1}{2} {u^2(x,t)} - \gamma 
\frac{\partial u(x,t)}{\partial x}
\end{equation}

\noindent
where $\gamma$ is a constant. Expanding $J$ again upto first order in $\tau$ 
and differentiating the resulting equation with respect to $x$ followed by 
differentiation of Eq.(2) for $k=0$ (i.e., in the absence of any source term)
with respect to time $t$ and elimination of $J$ as done in the last section 
we obtain 

\begin{equation}
\label{eq36}
\frac{\partial u}{\partial t} + u \frac{\partial u}{\partial x}
- \tau \frac{{\partial}^2 u}{\partial t^2}
= \gamma \frac{{\partial}^2 u}{\partial x^2} \; \; .
\end{equation}

\noindent
For $\tau$ = 0 Eq.(36) assumes the form of classical Burgers' equation 
\cite{ld,jmb} when
$u(x,t)$ and $\gamma$ are identified as the velocity field and kinematic 
viscosity, respectively.

We now seek a traveling wave solution of the Burgers' equation with memory
(36) in the form, $U(z) = u(x-ct)$, $z = x-ct$, where $c$ is again the wave
speed to be determined. This results in the following equation:

\begin{equation}
\label{eq37}
- \left ( \frac{c^2}{\beta} + \gamma \right ) 
\frac{{\partial}^2 U}{\partial z^2}
+ U \frac{\partial U}{\partial z}
- c \frac{\partial U}{\partial z} = 0
\end{equation}

\noindent
where $\beta = 1/ \tau$.

We now impose the bound condition on $U(z)$ that it asymptotically tends to
constant values $u_1$ as $z \rightarrow -\infty$ and $u_2$ as 
$z \rightarrow +\infty$ and $u_1 > u_2$.

A direct integration of (37) yields 
 
\begin{equation}
\label{eq38}
\frac{\partial U}{\partial z} = \frac{1}{2 (\frac{c^2}{\beta} + \gamma)}
\left (U^2 - 2 c U - 2 A \right )
\end{equation}

\noindent
where $A$ is the integration constant. If $u_1$ and $u_2$ are the roots of 
the quadratic equation $U^2 - 2 c U - 2 A = 0$, then the wave speed $c$ and
the constant $A$ can be obtained as

\begin{equation}
\label{eq39}
c = \frac{u_1 + u_2}{2} \:\:\:\: {\rm and} \:\:\:\:\: A = - \frac{1}{2}
u_1 u_2 \; \; .
\end{equation}

\noindent
Eq.(38) can then be rewritten in the form 

\begin{equation}
\label{eq40}
2 \left  (\frac{c^2}{\beta} + \gamma \right ) \frac{\partial U}{\partial z}
= (U-u_1) (U-u_2)
\end{equation}

\noindent
to integrate to obtain finally 

\begin{equation}
\label{eq41}
U(z) = \frac{1}{2} (u_1 + u_2) - \frac{1}{2} (u_1 - u_2) 
\tanh \left [\frac{z}{4 \delta} \right ]
\end{equation}

\noindent
where $\delta$ is given by 

\begin{equation}
\label{eq42}
\delta = \left ( \frac{\frac{c^2}{\beta} + \gamma}{u_1 - u_2} \right )
\; \; .
\end{equation}

The above analysis shows that the shape of the wave form is 
not only affected by kinematic viscosity $\gamma$ by also by an additional
contribution $c^2/ \beta$ due to finite relaxation time $\tau (= 1/ \beta)$
such that $(c^2/ \beta) + \gamma$ behaves as effective kinematic viscosity. 
It is thus apparent that the balance between the steepening effect of the
convection as well as smoothening effect due to kinematic viscosity 
is enhanced by the presence of the wave speed dependent term
$c^2/ \beta$. Thus although wave speed $c \left [ (u_1 + u_2)/2 \right ]$
itself remain unaffected by the finite memory effect in contrast to our earlier
case of Fisher equation, transmission layer thickness '$\delta$' - which is a
measure of shock thickness, increases for higher speed $c$ and relaxation time
$\tau$. This implies that as the wave moves faster the shock smoothens out more
and more so that the speed dependence of thickness $\delta$ makes the dynamics
self-regulating in the problem of interaction between convection and
diffusion.

%\section{Conclusions}

\noindent
{\it Conclusions}:
The existence of relaxation or delay time is an important feature in
reaction-diffusion and convection-diffusion systems. In this paper we have
shown that two prototypical representatives of these systems a generalized
Fisher equation and Burgers' equation can be solved exactly for finite
arbitrary delay time using conventional methods. While the wave speed is
significantly modified in the Fisher problem for finite memory transport,
speed of the traveling wave in the corresponding Burgers' problem remains
unaffected, delay time being effective in smoothening out the shock-wave
nature of the traveling wave. We also establish numerically that for the 
reaction-diffusion system the strength of the reaction term must not exceed
a critical limit to allow travelling wave front solutions to exist for
appreciable memory or relaxation effect.
In view of the fact that the studies on
reaction-diffusion and convection-diffusion with finite memory transport have
been applied to forest fire \cite{vm4} and population growth models 
\cite{vm1}, Neolithic transitions \cite{jf}
and in several other areas under various approximate schemes
\cite{eeh,kph1,kph2,th,wh1,wh2,mhk,ga,sf,jms}, we
believe that the present exact solutions for the generalized Fisher and
Burgers' problem are very much pertinent in this context.

\begin{acknowledgments}
This work was supported by the Council of Scientific and Industrial 
Research (C.S.I.R.), Government of India.
\end{acknowledgments}


\begin{thebibliography}{99}

\bibitem{ld} L. Debnath, {\it Nonlinear Partial Differential Equations
for Scientists and Engineers} (Birkh\"auser, Boston, 1997).

\bibitem{ire} I.R. Epstein and J.A. Pojman, {\it An Introduction to
Nonlinear Chemical Dynamics: Oscillations, Waves, Patterns and Chaos}
(Oxford, New York, 1998).

\bibitem{nfb} N.F. Britton, {\it Reaction-Diffusion Equations and their
applications to Biology} (Academic, New York, 1986).

\bibitem{jdm} J.D. Murray, {\it Mathematical Biology}, 
Second, Corrected Edition (Springer, Berlin, 1993).

\bibitem{raf} R.A. Fisher, Ann. Eugenics {\bf 7}, 353 (1937).

\bibitem{jmb} J.M. Burgers, Adv. Appl. Mech. {\bf 1}, 171 (1948).

\bibitem{fort} J. Fort and V. M\'endez, Rep. Prog. Phys. {\bf 65}, 895 (2002).

\bibitem{gr1} Th. Gallay and G. Raugel, ZAMP {\bf 48}, 451 (1997).
 
\bibitem{gr2} Th. Gallay and G. Raugel, Preprint patt-sol/9809007;
Preprint patt-sol/9812007 .

\bibitem{eeh} E.E. Holmes, Am. Nat. {\bf 142}, 779 (1993).

\bibitem{kph1} K.P. Hadeler, 
Can. Appl. Math. Quart. {\bf 2}, 27 (1994).

\bibitem{kph2} K.P. Hadeler, in {\it Reaction Transport Systems in 
Mathematics Inspired by Biology}, edited by V. Capasso and O. Diekmann,
CIME Lectures, Florence (Springer-Verlag, Berlin, 1998).

\bibitem{th} T. Hillen,
Math. Models Methods Appl. Sci. {\bf 8}, 507 (1998).

\bibitem{vm1} V. M\'endez and J. Camacho,
Phys. Rev. E {\bf 55}, 6476 (1997).

\bibitem{wh1} W. Horsthemke, Phys. Lett. A {\bf 263}, 285 (1999).

\bibitem{wh2} W. Horsthemke, Phys. Rev. E {\bf 60}, 2651 (1999).

\bibitem{mhk} K.K. Manne, A.J. Hurd and V.M. Kenkre,
Phys. Rev. E {\bf 61}, 4177 (2000).

\bibitem{ga} G. Abramson, A.R. Bishop and V.M. Kenkre,
Phys. Rev. E {\bf 64}, 066615 (2001).

\bibitem{sf} S. Fedotov, Phys. Rev. Lett. {\bf 86}, 926 (2001).

\bibitem{jms} J.M. Sancho and A. S\'anchez, Phys. Rev. E
{\bf 63}, 056608 (2001).

\bibitem{vm4} V. M\'endez and J.E. Llebot,
Phys. Rev. E {\bf 56}, 6557 (1997).

\bibitem{jf} J. Fort and V. M\'endez,
Phys. Rev. Lett. {\bf 82}, 867 (1999).

\bibitem{kbr} S. Kar, S.K. Banik and D.S. Ray, 
Phys. Rev. E {\bf 65}, 061909 (2002).

\bibitem{cc} C. Cattaneo, C. R. Acad. Sci. {\bf 247}, 431 (1958).

\bibitem{mkot} M. Kot, {\it Elements of Mathematical Ecology}
(Cambridge University Press, Cambridge, 2001).

\end{thebibliography}
\end{document}